\documentclass[12pt,thmsa]{article}
\usepackage{amsfonts}
\usepackage{amssymb}
\usepackage{graphicx}

\input{tcilatex}
\begin{document}

\begin{center}
{\Large Quantum deficits and correlations of two superconducting charge
qubits}

{\small {N. Metwally and M. Abdel-Aty} }

{\small {\footnotesize Mathematics Department, College of Science, Bahrain
University, 32038 Bahrain \\[0pt]
}}
\end{center}

Motivated by recent experiments [Pashkin et al. Nature, \textbf{421}, 823
(2003)] which showed coherent oscillations of two superconducting qubits
system, we consider a system of two charge qubits coupled to a common
stripline microwave resonator. We discuss the separable and entangled
behavior as well as the quantum and classical information deficits.
Numerical computation of these quantities for several regimes is performed.
We find that for less entangled states the partner can extract much more
information by means of classical communication and local operations.\bigskip

\section{Introduction}

Fundamental quantum phenomena, such as non-locality and entanglement of
quantum degrees of freedom, have regained a lot of interest recently, mainly
due to their potential usefulness as a computational resource. There is a
considerable interest in the study of the quantum deficits and quantum
correlations properties of systems that are comprised of qubits, with the
aim of using such systems for quantum information processing \cite%
{mew07,pas03}. An understanding of the correlations properties of such a
system is of great use since it provides information about coherence and the
transmission channels in a multi-qubit system. Such information could be
useful for the study of controlled, decoherence-free information transfer
through the channel.

On the other hand, the realization of a solid-state qubit based on familiar
and highly developed semiconductor technology would facilitate scaling to a
many-qubit computer and make quantum computation more accessible \cite{hen07}%
. An attractive point of that proposal is the large spin decoherence time
characteristic of semiconductors; a drawback is that it requires local
control of intense magnetic fields. As an alternative, a spin-based logical
qubit involving a multiple charge qubits setup and voltage controlled
exchange interactions was devised, but at the price of considerable overhead
in additional operations \cite{aws02,los98}. Also, entangled qubits are very
important in the context of achieving quantum teleportation \cite{Riebe,
Hammer, Prry}, quantum coding \cite{Metwally, Wang} and cryptography \cite%
{ben84}. Recently, investigation of classical, quantum and total correlation
has been attracting many authors. For example, Prants et. al. \cite{Prants}
have investigated the stability and instability of quantum evolution in the
interaction between a 2-two level atom with recoil photon and a quantized
field mode in an ideal cavity. X. Cui et. al. have used the correlation
entropy to quantify the correlation between two qubits \cite{Cui}.
Quantifying the quantum and classical deficits by using the quantum and
classical correlations are investigated by Pankowski et al. \cite{Pank}. A
quantum correlation of a physically interesting system interacting with its
environment in the context of two-coupled superconducting charges model has
been\ discussed in Ref. \cite{aty}.

In this paper we consider a class of a two-qubit system consists of two
superconducting charge qubits interact with a common stripline microwave
resonator. We carefully investigate the effective dynamics of the qubits
subsystem in a regime where the two identical superconducting charge qubits
coupled capacitively to a stripline resonator. The behavior of classical and
quantum correlations are discussed for different regimes. Also this depends,
amongst other things, on the initial state settings. Finally, we evaluate
the quantum and classical deficits for a generated entangled state between
the two qubits.

\section{Two qubits coupled to a stripline resonator}

Here we consider two identical superconducting charge qubits each of which
is coupled capacitively to a stripline resonator. If the qubits are placed
at an antinode of the fundamental harmonic mode of the resonator, then we
can describe the system as a pair of two-level systems coupled to a simple
harmonic oscillator. The charging energy of the qubits and their coupling to
the resonator can be controlled by the application of magnetic and electric
fields \cite{nak99,rod08}. If these are tuned so that the qubits are close
to resonance with the field, then by using the rotating wave approximation,
the the system can be described by,

\begin{equation}
\hat{H}_{in}=\sum_{j=1}^{2}\left\{ \Gamma {_{1}\sigma _{z}^{(j)}+}\lambda
(\psi ^{\dagger }\sigma _{-}^{(i)}+h.c.)\right\} ,
\end{equation}%
where $\psi$ and $\psi ^{\dagger }$ are the annihilation and creation
operators of the photons, $\Gamma _{j}$ the energy of the $j^{th}$ Cooper
pair box,$\lambda $ the resonator-qubit coupling and $\sigma ^{\prime }s$
are the Pauli matrices.

In the invariant sub-space of the global system, we can consider a set of
complete basis of the qubit-field system as $\bigl|ee,n\bigr\rangle,\bigl|%
eg,n+1\bigr\rangle,\bigl|ge,n+1\bigr\rangle$ and $\bigl|gg,n+2\bigr\rangle$.
The time evolution of the density operator of the system is given by 
\begin{equation}
\varrho _{cf}(t)=\mathcal{U}(t)\{\varrho _{a}(0)\otimes \varrho _{f}(0)\}%
\mathcal{U}^{\dagger }(t),  \label{Denst}
\end{equation}%
where $\mathcal{U}(t)=\exp \left( -i\hat{H}t/\bar{h}\right) $ is the unitary
operator, its components are given by, 
\begin{eqnarray}
\mathcal{U}_{11}(t) &=&\sum_{i=1}^{3}{(-1)^{i+1}\alpha _{i}e^{-i\mu
_{1}t}[\mu _{i}(\Delta +\mu _{i})-2\beta ^{2}]},  \nonumber \\
\mathcal{U}_{12}(t) &=&\gamma \sum_{i=1}^{3}{(-1)^{i+1}e^{-i\mu
_{i}t}(\Delta +\mu _{i})},  \nonumber \\
\mathcal{U}_{13}(t) &=&\mathcal{U}_{12}(t),\quad \mathcal{U}_{14}(t)=2\beta
\gamma \sum_{i=1}^{3}{(-1)^{i+1}\alpha _{i}e^{-i\mu _{i}t}},  \nonumber \\
\mathcal{U}_{22}(t) &=&\sum_{i=1}^{3}{(-1)^{i+1}\frac{\alpha _{i}}{\mu _{i}}%
e^{-i\mu _{i}t}\Bigl[(\beta ^{2}(\Delta -\mu _{i})-(\Delta +\mu
_{i}))(\gamma ^{2}+\mu _{i}(\Delta -\mu _{i}))\Bigr]}  \nonumber \\
&&-\frac{\Delta (\beta ^{2}-\gamma ^{2})}{\mu _{1}\mu _{2}\mu _{3}}, 
\nonumber \\
\mathcal{U}_{23}(t) &=&-\sum_{i=1}^{3}{(-1)^{i+1}\frac{\alpha _{i}}{\mu _{i}}%
e^{-i\mu _{i}t}\Bigl[\beta ^{2}(\Delta -\mu _{i})-\gamma ^{2}(\Delta +\mu
_{i})\Bigr]}+\frac{\Delta (\beta ^{2}-\gamma ^{2})}{\mu _{1}\mu _{2}\mu _{3}}%
,  \nonumber \\
\mathcal{U}_{24}(t) &=&-\beta \sum_{i=1}^{3}{(-1)^{i+1}\alpha _{i}e^{-i\mu
_{i}t}(\Delta -\mu _{i})},  \nonumber \\
\mathcal{U}_{31}(t) &=&\mathcal{U}_{13}(t),\quad \mathcal{U}_{32}(t)=%
\mathcal{U}_{23}(t),\quad \mathcal{U}_{33}(t)=\mathcal{U}_{22}(t),  \nonumber
\\
\mathcal{U}_{41}(t) &=&\mathcal{U}_{14}(t),\quad \mathcal{U}_{42}(t)=%
\mathcal{U}_{24}(t),\quad \mathcal{U}_{43}(t)=\mathcal{U}_{34}(t),  \nonumber
\\
\mathcal{U}_{44}(t) &=&-\sum_{i=1}^{3}{(-1)^{i+1}\alpha _{i}e^{-i\mu _{i}t}%
\Bigl[2\gamma ^{2}+\mu _{i}(\Delta -\mu _{i})\Bigr]},
\end{eqnarray}%
where $\gamma =\sqrt{n+1},\beta =\sqrt{n+2}$,$\alpha _{1}=(\mu _{12}\mu
_{13})^{-1},~\alpha _{2}=(\mu _{12}\mu _{23})^{-1},\alpha _{3}=(\mu _{13}\mu
_{23})^{-1}$, $\mu _{kj}=\mu _{k}-\mu _{j}$ and $\mu _{i}=\frac{2}{3}\kappa
\cos \theta _{i}$ with $\kappa =\sqrt{3(\Delta ^{2}+2(\beta ^{2}+\gamma
^{2}))}$ and $\theta _{1}=\frac{1}{3}\cos ^{-1}\left( -\frac{27\Delta }{%
\kappa ^{3}}\right) $,~ $\theta _{2}=\frac{2\pi }{3}+\theta _{1},~\theta
_{3}=\frac{2\pi }{3}+\theta _{2}$.

Since we are interested in discussing some properties of the charge qubits
system, we calculate the density matrix of the charged qubit by tracing out
the field i.e $\varrho _{ab}=tr_{f}\{\varrho _{cf}\}$ . 
\begin{eqnarray}
\varrho _{ab}(t) &=&A_{n}^{(1)}|^{2}\bigl|gg\bigr\rangle\bigl\langle gg\bigr|%
+A_{n}^{\ast (1)}A_{n+2}^{(2)}\bigl|eg\bigr\rangle\bigl\langle gg\bigr|%
+A_{n}^{\ast (1)}A_{n+1}^{(3)}\bigl|ge\bigr\rangle\bigl\langle gg\bigr| 
\nonumber \\
&&+A_{n}^{\ast (1)}A_{n+2}^{(4)}\bigl|ee\bigr\rangle\bigl\langle gg\bigr|%
+A_{n}^{(1)}A_{n+2}^{\ast (2)}\bigl|gg\bigr\rangle\bigl\langle eg\bigr|%
+|A_{n}^{(2)}|^{2}\bigl|eg\bigr\rangle\bigl\langle eg\bigr|  \nonumber \\
&&+A_{n}^{\ast (2)}A_{n}^{(3)}\bigl|ge\bigr\rangle\bigl\langle eg\bigr|%
+A_{n}^{\ast (2)}A_{n+1}^{(4)}\bigl|ee\bigr\rangle\bigl\langle eg\bigr|%
+A_{n}^{(1)}A_{n+1}^{\ast (3)}\bigl|gg\bigr\rangle\bigl\langle ge\bigr| 
\nonumber \\
&&+A_{n}^{(2)}A_{n}^{\ast (3)}\bigl|eg\bigr\rangle\bigl\langle ge\bigr|%
+|A_{n}^{(3)}|^{2}\bigl|ge\bigr\rangle\bigl\langle ge\bigr|+A_{n}^{\ast
(3)}A_{n+1}^{(4)}\bigl|ee\bigr\rangle\bigl\langle ge\bigr|  \nonumber \\
&+&A_{n}^{(1)}A_{n+2}^{\ast (4)}\bigl|gg\bigr\rangle\bigl\langle ee\bigr|%
+A_{n}^{\ast (2)}A_{n+1}^{(4)}\bigl|eg\bigr\rangle\bigl\langle ee\bigr| 
\nonumber \\
&&+A_{n}^{(3)}A_{n+1}^{\ast (4)}\bigl|ge\bigr\rangle\bigl\langle ee\bigr|%
+|A_{n}^{(4)}|^{2}\bigl|ee\bigr\rangle\bigl\langle ee\bigr|,
\end{eqnarray}%
with, 
\begin{eqnarray}
A_{n}^{(1)} &=&\sum_{j=1}^{4}{\mathcal{U}_{1j}(n)A^{(j)}(0)},\quad
A_{n}^{(2)}=\sum_{j=1}^{4}{\mathcal{U}_{2j}(n)A^{(j)}(0)},\quad   \nonumber
\\
A_{n}^{(3)} &=&\sum_{j=1}^{4}{\mathcal{U}_{3j}(n)A^{(j)}(0)},\quad
A_{n}^{(4)}=\sum_{j=1}^{4}{\mathcal{U}_{4j}(n)A^{(j)}(0)},
\end{eqnarray}%
where, $%
A^{(1)}(0)=b_{1}b_{2},~A^{(2)}(0)=b_{1}a_{2},~A^{(3)}(0)=a_{1}b_{2},~A^{(4)}(0)=a_{1}a_{2}.
$

\section{Correlation dynamics}

The correlations in a state $\rho _{ab}$ can be split into two parts,
quantum and classical parts \cite{Vedral}. The classical part is seen as the
amount of information bout one subsystem, say $A$, that can be obtained by
performing a measurement on the other subsystem, $B$. The total correlation $%
\mathcal{T}_{c}$ can be measured by the Von Neumann mutual information
between the two subsystems, which is defined by 
\begin{equation}
\mathcal{T}_{c}=\mathcal{S}(\varrho _{a})+\mathcal{S}(\varrho
_{b})-S(\varrho _{ab}),
\end{equation}%
where $\mathcal{S}(i)=-tr{\varrho _{i}log\varrho _{i}}$, $i=a,b$ and $ab$.
Bipartite system consists of two states $\varrho _{a}$ and $\varrho _{b}$ is
called classical correlated if one can write the total state as a tensor
product of the individual states, i.e $\varrho _{ab}=\varrho _{a}\otimes
\varrho _{b}$. These types of systems are called separable, which has no
entanglement. So that the total correlations comes from the classical
correlation \cite{Pop, Wen, Shi}. If we cannot write the total state as a
tensor product of its individual subsystems, then the states are quantum
correlated. These types of correlation can be evaluated by calculating the
amount of entanglement contained in the system. For bipartite systems, there
are several measures to quantify the degree of entanglement. Among these
measures, concurrence, entanglement of formation \cite{Hill} and negativity 
\cite{Zyc}. The negativity measure states that if the eigenvalues of the
partial transpose of the total state $\varrho _{ab}$ are given by $\lambda
_{\xi },\xi =1,2,3,4$, then the degree of entanglement is, 
\begin{equation}
\mathcal{Q}_{c}=\sum_{\xi =1}^{4}{|\lambda _{\xi }|-1}.
\end{equation}%
For maximally entangled state the degree of entanglement is unity, while for
mixed entangled states $0\leq \mathcal{Q}_{c}<1$. Physically, the quantum
correlation measures the amount of immediate effect on one subsystem say, $%
\varrho _{a(b)}$ when we measure the other system $\varrho _{b(a)}$ \cite%
{Wen}. On the other hand, the quantum correlation is fragile, so due to the
decoherence a maximally entangled state can be turned into a completely
separable state. On other words the quantum correlation can be completely
erased, while the classical correlation cannot be erased \cite{Gogo}. So the
classical correlation can be defined as the difference between the total
correlation and the quantum correlation\cite{Wen, Pank}, 
\begin{equation}
\mathcal{C}_{c}=\mathcal{T}_{c}-\mathcal{Q}_{c}
\end{equation}%
\begin{figure}[t]
\begin{center}
\includegraphics[width=20pc,height=12pc]{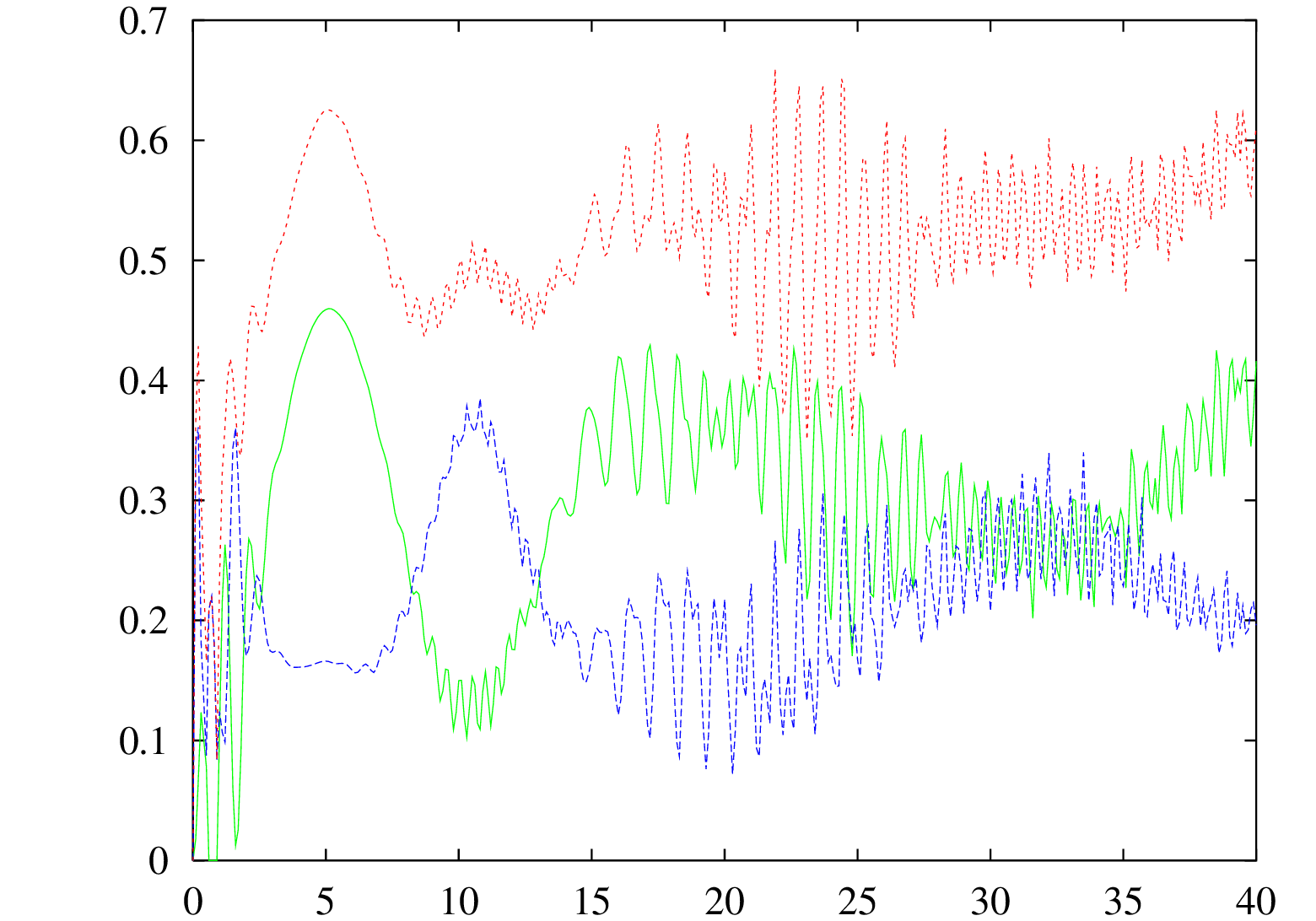}\
\put(-200,125){(a)}\ \put(-105,-10){ $\lambda t$} %
\includegraphics[width=20pc,height=12pc]{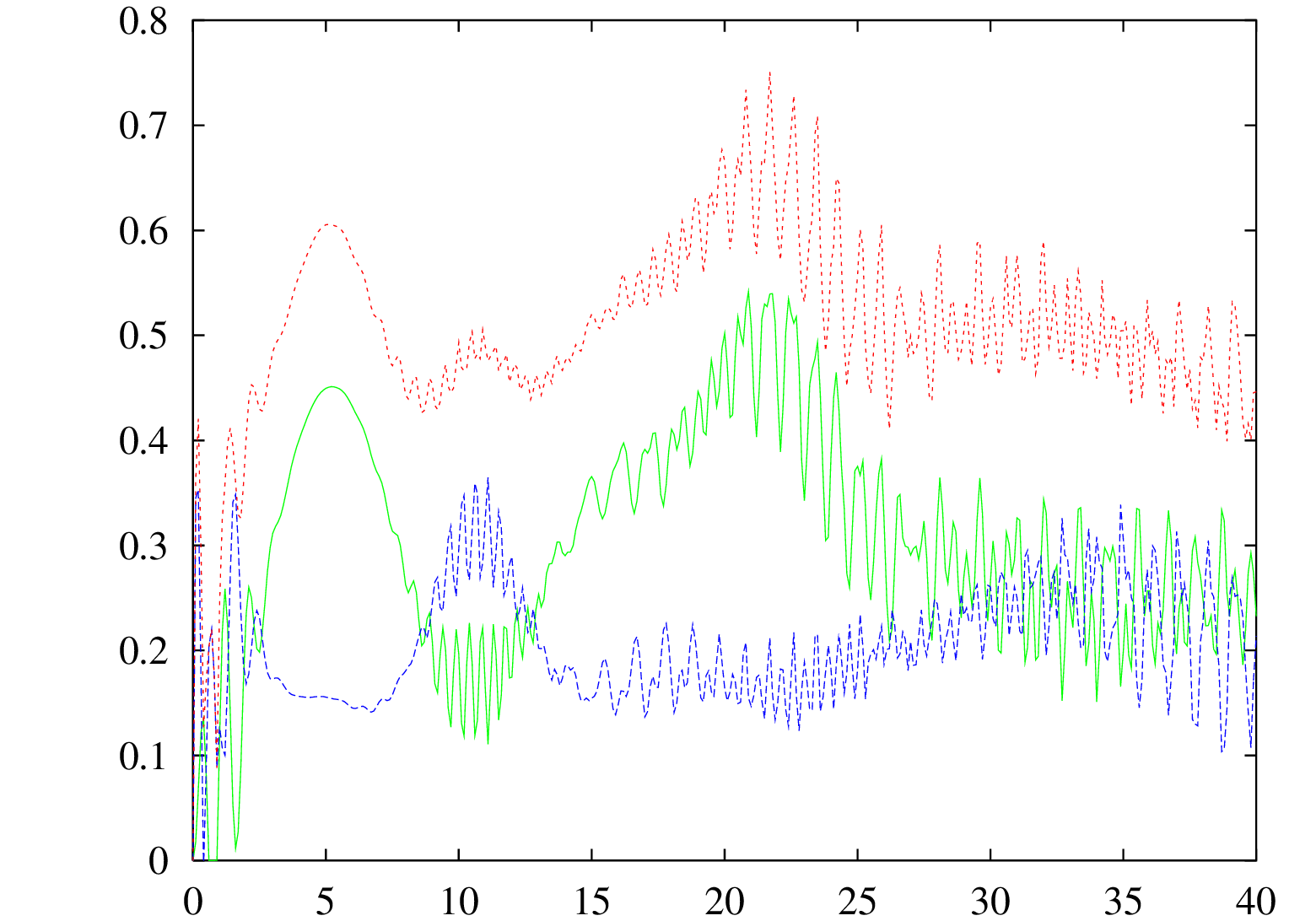} \put(-200,125){(b)}\
\put(-105,-10){ $\lambda t$} \ %
\includegraphics[width=20pc,height=12pc]{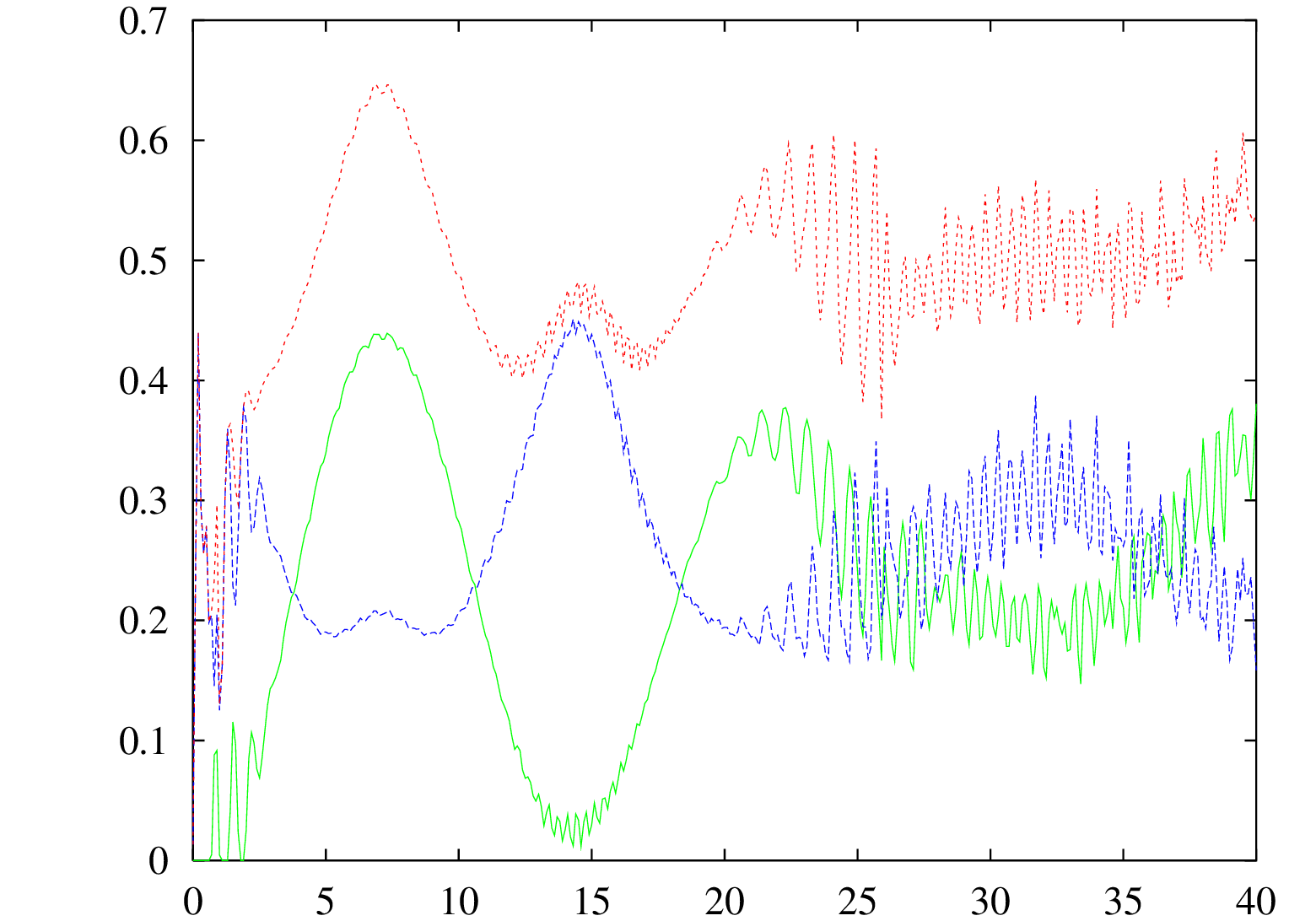}~
\put(-200,125){(c)}\ \put(-105,-10){ $\lambda t$} %
\includegraphics[width=20pc,height=12pc]{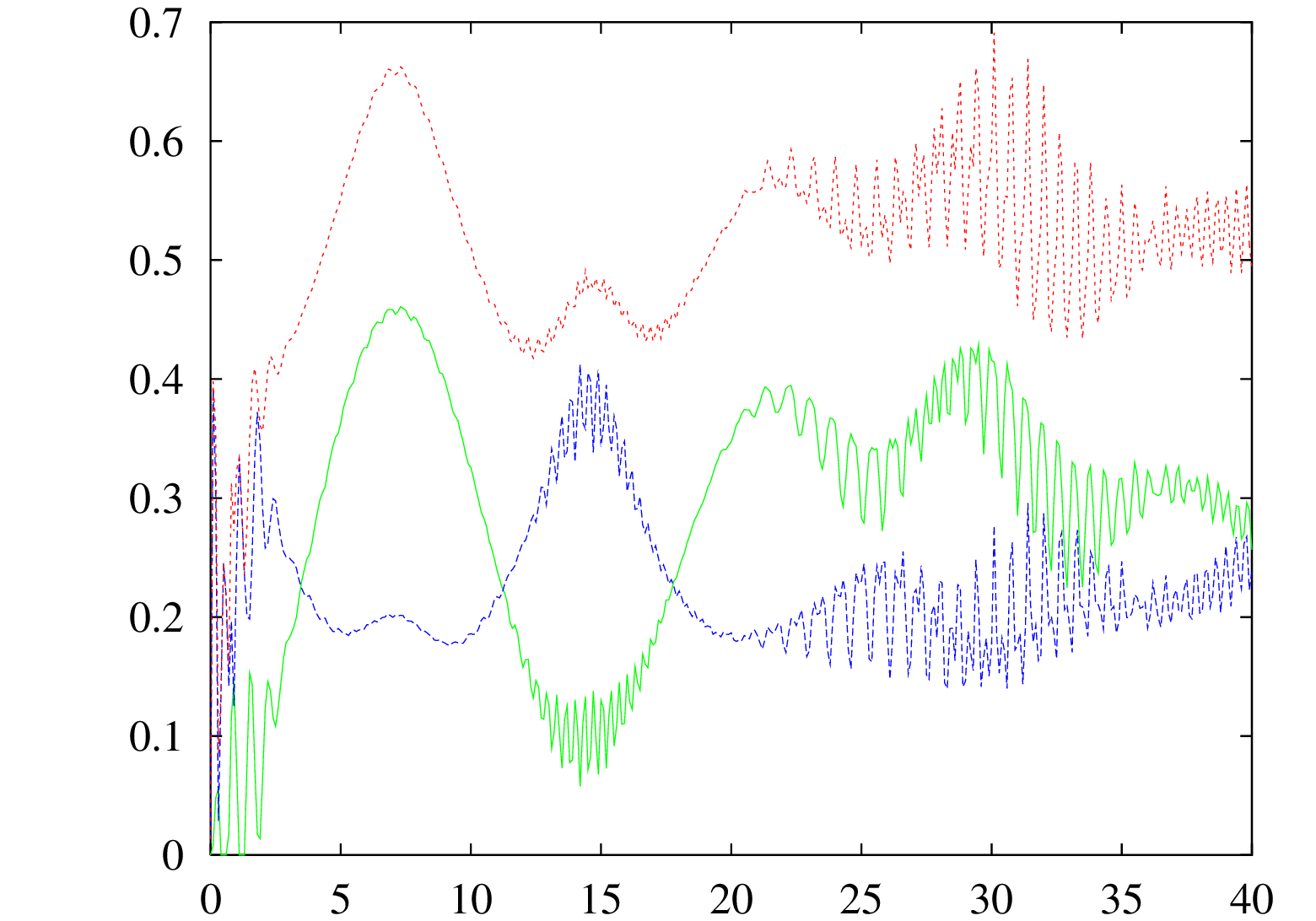} \put(-200,125){(d)}\
\put(-105,-10){ $\lambda t$}
\end{center}
\caption{ The total correlation for the system in excited state (a)$\frac{%
\Delta }{\protect\lambda }=0.5$ (b) $\frac{\Delta }{\protect\lambda }=1.0$
where $\bar{n}=10$.(C)$\frac{\Delta }{\protect\lambda }=0.5 $(d) $\frac{%
\Delta }{\protect\lambda }=1$ where $\bar{n}=20$.}
\end{figure}
In Fig.(1), we assume that the qubits system is initially prepared in
excited state i.e $\rho _{ab}(0)=|ee\rangle \langle ee|$. The effect of the
detuning parameter $\Delta $ on the dynamics of correlation is shown in Figs.%
$(1a)$ and $(1b)$, where we plot the quantum, classical and the total
correlations. It is clear that at $t=0$ the system is completely separable,
so there is no any correlations between the two qubits. For a short period
of time there is no quantum correlation but the classical and total
correlations take place in a similar manner, $\mathcal{T}_{c}=\mathcal{C}%
_{c} $. As time evolves the quantum correlation appears and increases on the
expanse of the classical correlation. At a certain range of time, the
classical correlation exceeds the quantum correlation i. e $\mathcal{C}_{c}>%
\mathcal{Q}_{c}$, while for some other ranges $\mathcal{Q}_{c}>\mathcal{C}%
_{c}$. From theses figures we can see that for small values of the detuning
parameter the quantum correlation tends to a minimum value while the
classical correlation reaches to a maximum value at $\lambda t\simeq 10$ for
Fig. $(1a)$.

The situation is changed considerably when the detuning is increased, where
at the corresponding point which has been observed in Fig. $(1a)$, the
minimum values of the quantum correlation increases on the expanse of the
classical correlation. Also, the maximum values of $\mathcal{Q}_c$ as well
as the intervals in which $\mathcal{C}_c>\mathcal{Q}_c$ decreases as the
detuning parameter increases. These behaviors are seen by comparing Figs.($%
1a)$ and $(1b)$. The same remarks are seen in Figs.$(1c)$ and $(1d) $, where
we consider $\bar n=20$. This means that as $\frac{\Delta}{\lambda}$ is
increased a strong quantum correlation obtained, where the intervals of
times in which $\mathcal{Q}_c> \mathcal{C}_c$ increase for large values of
the detuning parameter So one can consider the detuning parameter as a
control parameter to generate a long lived entanglement, where the two
qubits entangled forever \cite{aty}.

The effect of the mean photon number $\bar{n}$ is shown by comparing Figs. $%
(a,b)$ and Figs. $(c,d)$, where we assume that $\bar{n}=10,20$ respectively.
From these figures the usual Rabi oscillation is seen, where the location of
its oscillation is moved toward the positive direction as the time
increases. Also, as the mean photon number increases the oscillations
decreases and hence both classical and quantum correlations increase. This
is clear by comparing Figs.$(1a)$ where we consider $\bar n=10$ and Fig.$%
(1c) $, where $\bar n=20$. We can see that although the maximum values of
the $\mathcal{Q}_c$ decreases as $\bar n$ increases, the quantum
correlations are more stable. This phenomena is seen clearly by comparing
Figs.$(1b)$ and $(1d)$

\begin{figure}[tbp]
\begin{center}
\includegraphics[width=20pc,height=12pc]{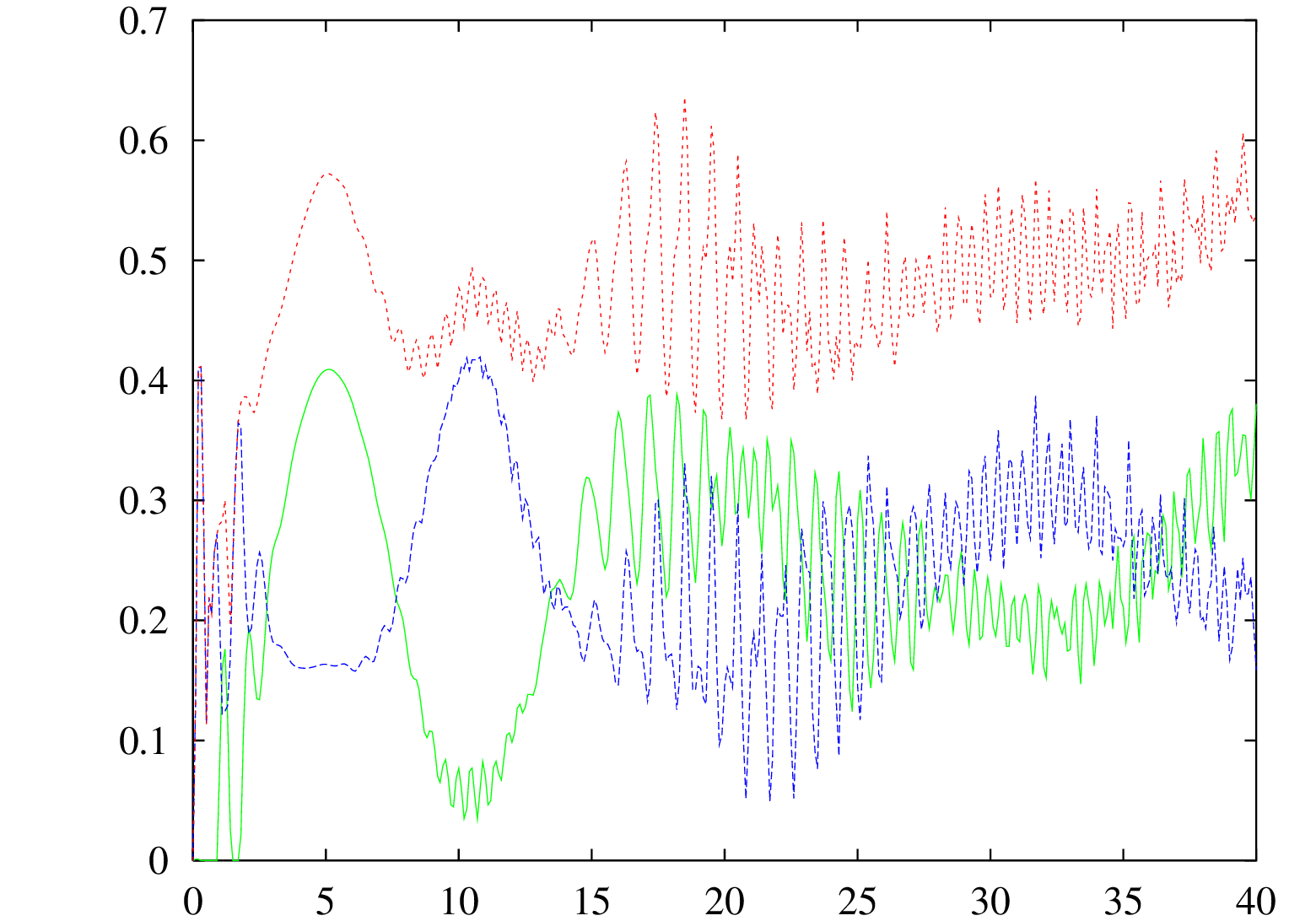}\put(-200,125){(a)}\
\put(-105,-10){ $\lambda t$} %
\includegraphics[width=20pc,height=12pc]{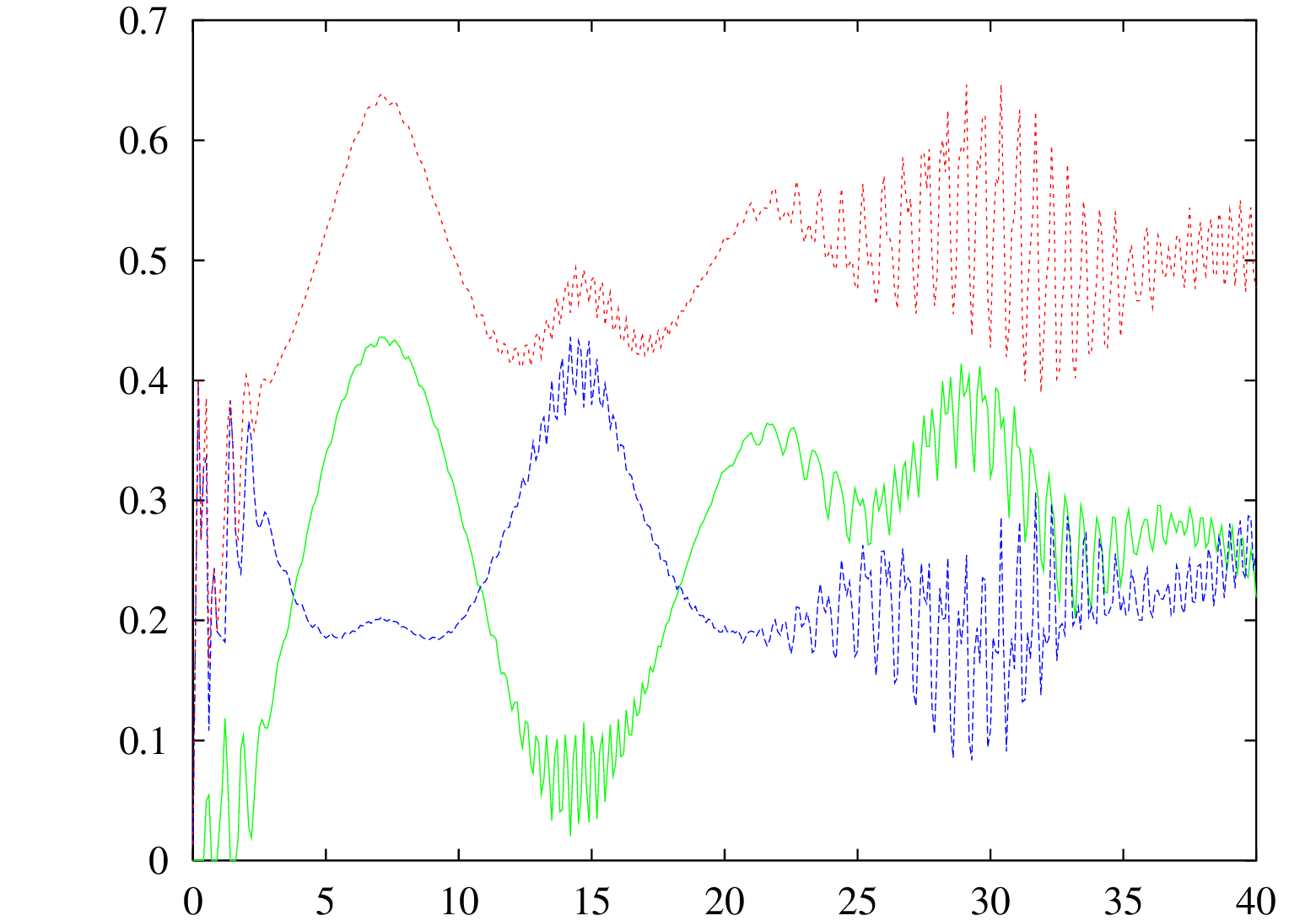}\
\put(-200,125){(b)}\ \put(-105,-10){ $\lambda t$}
\end{center}
\caption{ The total correlation for a system in ground state (a)$\frac{%
\Delta }{\protect\lambda }=0.5$ and $\bar{n}=10$ (b) $\frac{\Delta }{\protect%
\lambda }=1.0$ and $\bar{n}=20$}
\end{figure}

In Fig.$(2)$, we assume that the initial state of the atomic system is
prepared in the ground state, $\psi _{AB}(0)=\bigl|gg\bigr\rangle$. By
comparing the classical and quantum correlations in Fig. $(1a)$ and Fig. $%
(2a)$, one can see that the minimum value of the quantum correlation round $%
\lambda t=10$, in Fig.$(1a)$ is larger than that depicted on its
corresponding at Fig.$(2a)$. Also, the intervals of times in which the
quantum correlation is larger than the classical increases for the atomic
system is prepared initially in the excited state Also, the same remark is
seen in Fig. $(1d)$ and Fig. $(2b)$, where the minimum value of the quantum
correlation in the later figure is smaller than that for Fig$.(1d)$.

From the preceding results, we can say that the quantum and classical
correlations are sensitive for the detuning parameter, mean photon number
and the initial state setting of the atomic system. One can recruitment
these parameter to generate long-lived entangled state, where starting by
atomic system in excited state, large detuning parameter and small mean
photon number is one of the best choice to generate a useful entangled state
for quantum information tasks. 
\begin{figure}[tbp]
\begin{center}
\includegraphics[width=18pc,height=12pc]{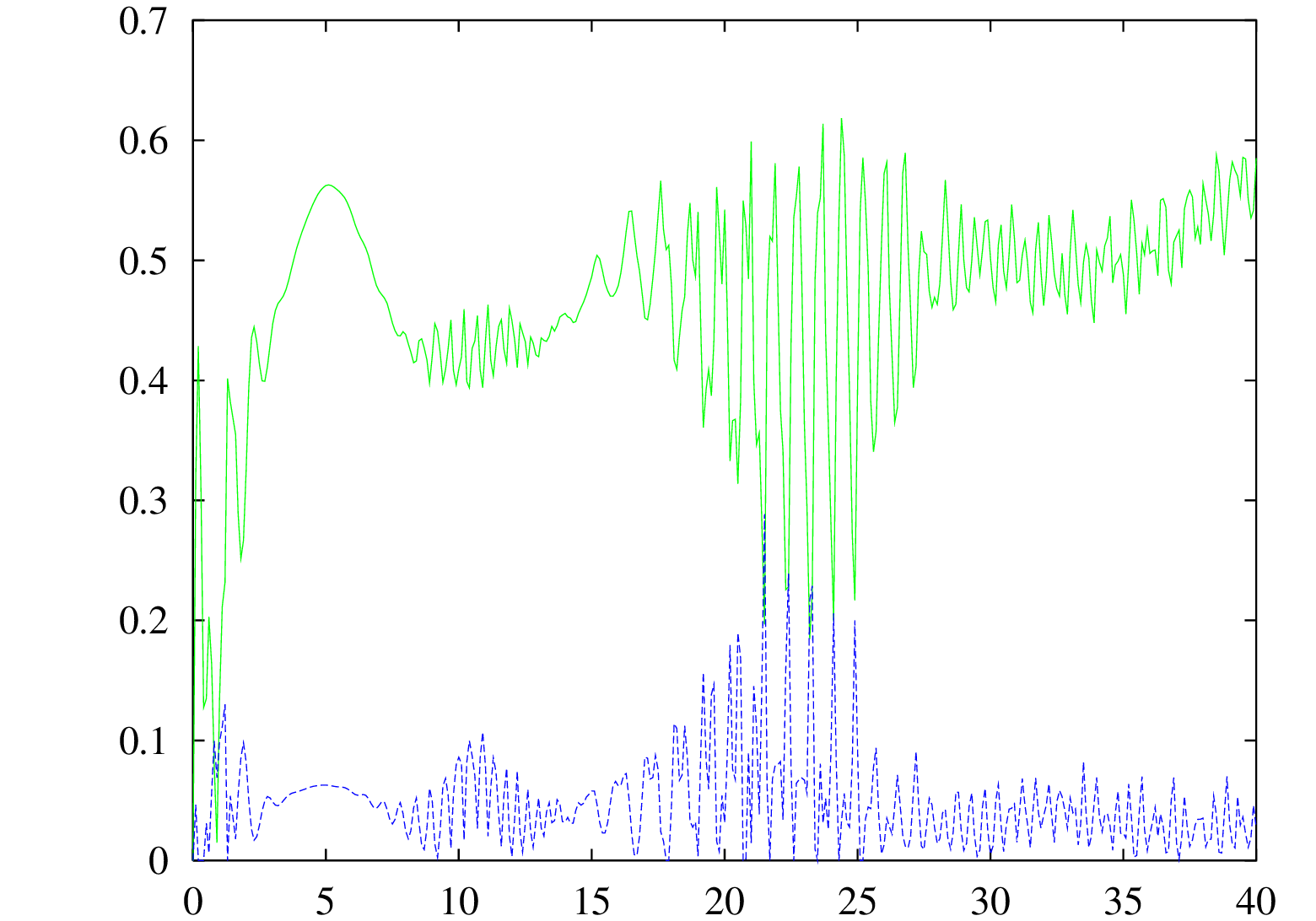} \put(-180,130){ (a)
} \put(-105,-10){ $\lambda t$} %
\includegraphics[width=18pc,height=12pc]{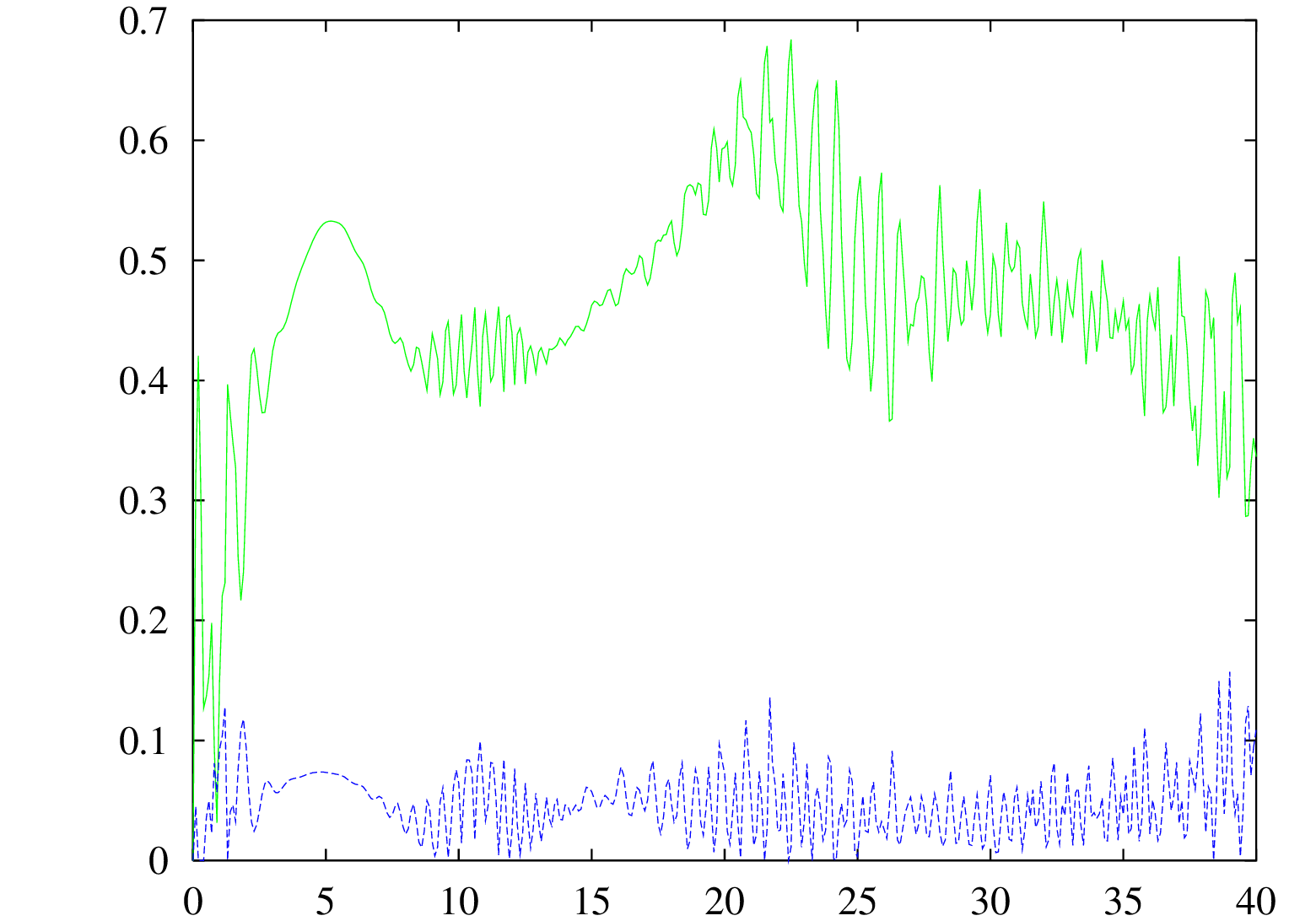} \put(-180,130){ (b)
} \put(-105,-10){ $\lambda t$}
\end{center}
\caption{ The solid and dot curves represent the quantum and classical
deficits for a system initially prepared in an excited state, where $\bar
n=10$ and $\frac{\Delta}{\protect\lambda}=0.5,1$ for Fig.(a) and Fig.(b)
respectively. }
\end{figure}

\section{Deficits}

To measure how much of correlations that must be destroyed during the
process of localizing information into a subsystem, we have to evaluate what
is called quantum deficit. It is defined as the difference between the
informational content in a state and information that can be localize to a
subsystem by the use of local unitary operations and a dephasing channel 
\cite{Hord}. The localizable information of a state $\varrho _{ab}$ is
defined by the maximal amount of local information. In a mathematical form
it is defined as, 
\begin{equation}
\mathcal{I}_{loz}=Sup_{\Lambda \in LOCC}(\mathcal{I}(\varrho _{a})+\mathcal{I%
}(\varrho _{b})),
\end{equation}%
under the local operation and classical communication, LOCC. Then we can
define the quantum deficit as 
\begin{equation}
\mathcal{Q}_{def}=\mathcal{I}({\varrho _{AB}})-\mathcal{I}_{loz}.
\end{equation}%
Also in this context, it is important to evaluate the classical information
deficit, $\mathcal{C}_{def}$. To achieve this aim we evaluate the local
information, the total information which contained in the individual
subsystems $\varrho _{a}$ and $\varrho _{b}$, $\mathcal{I}_{Lo}$. This
quantity is defined by \cite{Hord}-\cite{MPR}. 
\begin{equation}
\mathcal{I}_{Lo}=\mathcal{I}(\varrho _{a})+\mathcal{I}(\varrho _{b})
\end{equation}%
Then the classical information deficit is, 
\begin{equation}
\mathcal{C}_{def}=\mathcal{I}_{Loz}-\mathcal{I}_{Lo}.
\end{equation}%
This quantity quantify the amount of information that can be obtained from
the state $\varrho _{ab}$ by exploiting additional correlations.

Given a specific initial state settings, it is known how to calculate the
actual quantum and classical deficits in the interaction channel. Our
investigation can be highlighted by a simple open question: what is the
difference between highest quantum and classical deficits that will be
obtained according different initial states? In Fig.$(3)$, we plot the
quantum and classical deficits for the charged qubits system initially
prepared in the excited state, $\bigl|\psi _{ab}(0)\bigr\rangle=\bigl|gg%
\bigr\rangle$ and the mean photon number $\bar{n}=10$, for different values
of the detuning parameter. From this figure, it is clear that $\mathcal{Q}%
_{def}$ is zero for separable states at $\tau =0$. As the quantum
correlation increase, the quantum deficit increases. This is because an
entangled state is generated with high degree of entanglement. As an example
at $\lambda t\simeq 5$, the generated entangled state has a high degree of
entanglement and hence the $\mathcal{Q}_{def}$ increases. Also we can see
that the quantum correlation is maximum round $\lambda t\simeq 20$ (see Fig.
1b). this means that an entangled state is generated with a high degree of
entanglement. If we take a look at the same point in Fig.(3b), it is simply
see that the quantum deficit is maximum, this is due to the weakness of the
entanglement. So, one can say that for entangled states which have high
degree of entanglement, the amount of information that destroyed in the
process of localizing is much larger than for that observed in the less
entangled states.

In Fig. $(4)$ we investigating a different setting of the charged qubit,
where we consider the qubits are initially prepared in the ground state. The
figure shows as one increases the detuning and the mean photon number the
quantum deficit increases while the classical deficit decreases. This
behavior due to the increasing of the degree of entanglement for these
parameters.

\begin{figure}[tbp]
\begin{center}
\includegraphics[width=18pc,height=12pc]{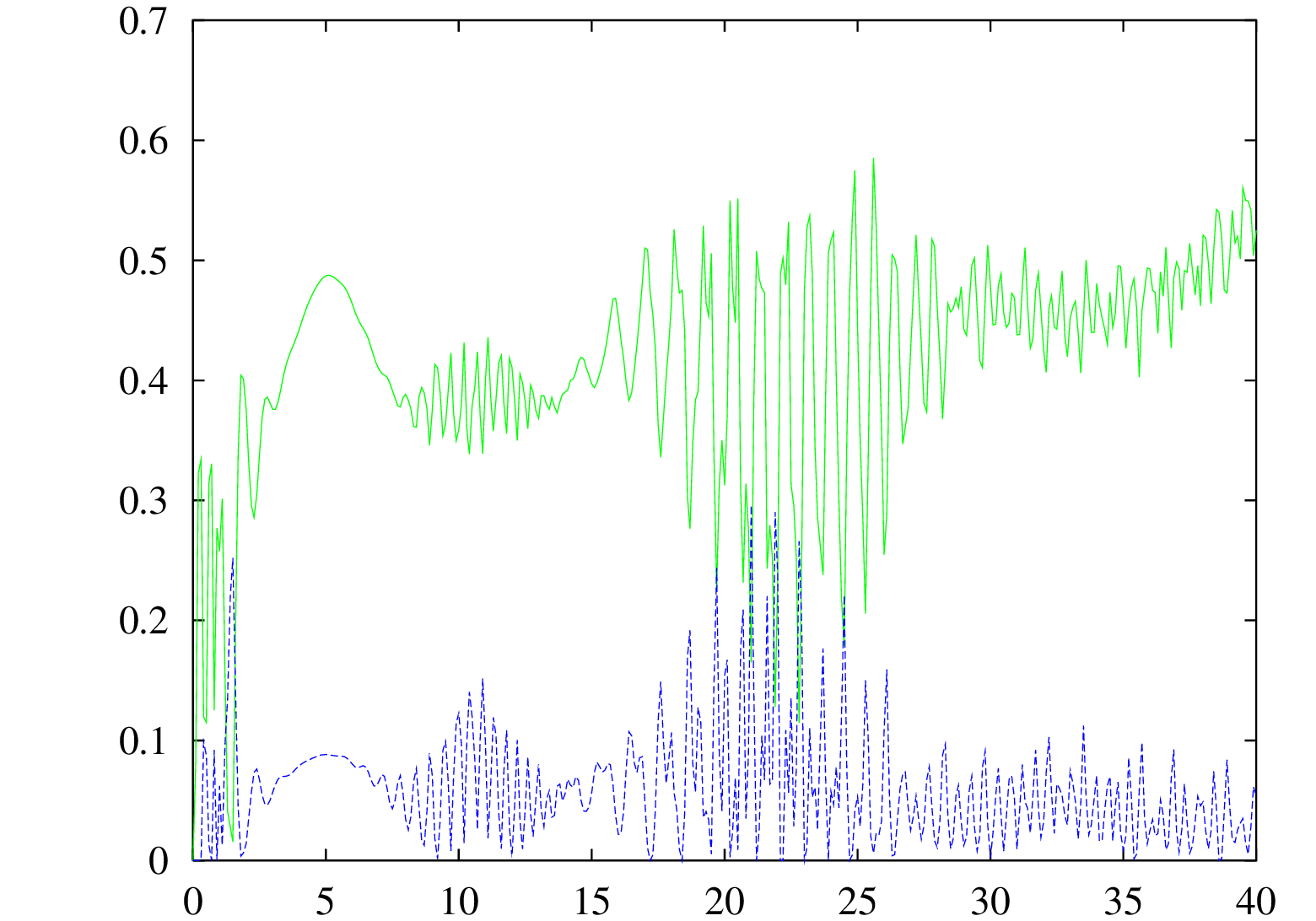} \put(-180,130){ (a)
} \put(-105,-10){ $\lambda t$} %
\includegraphics[width=18pc,height=12pc]{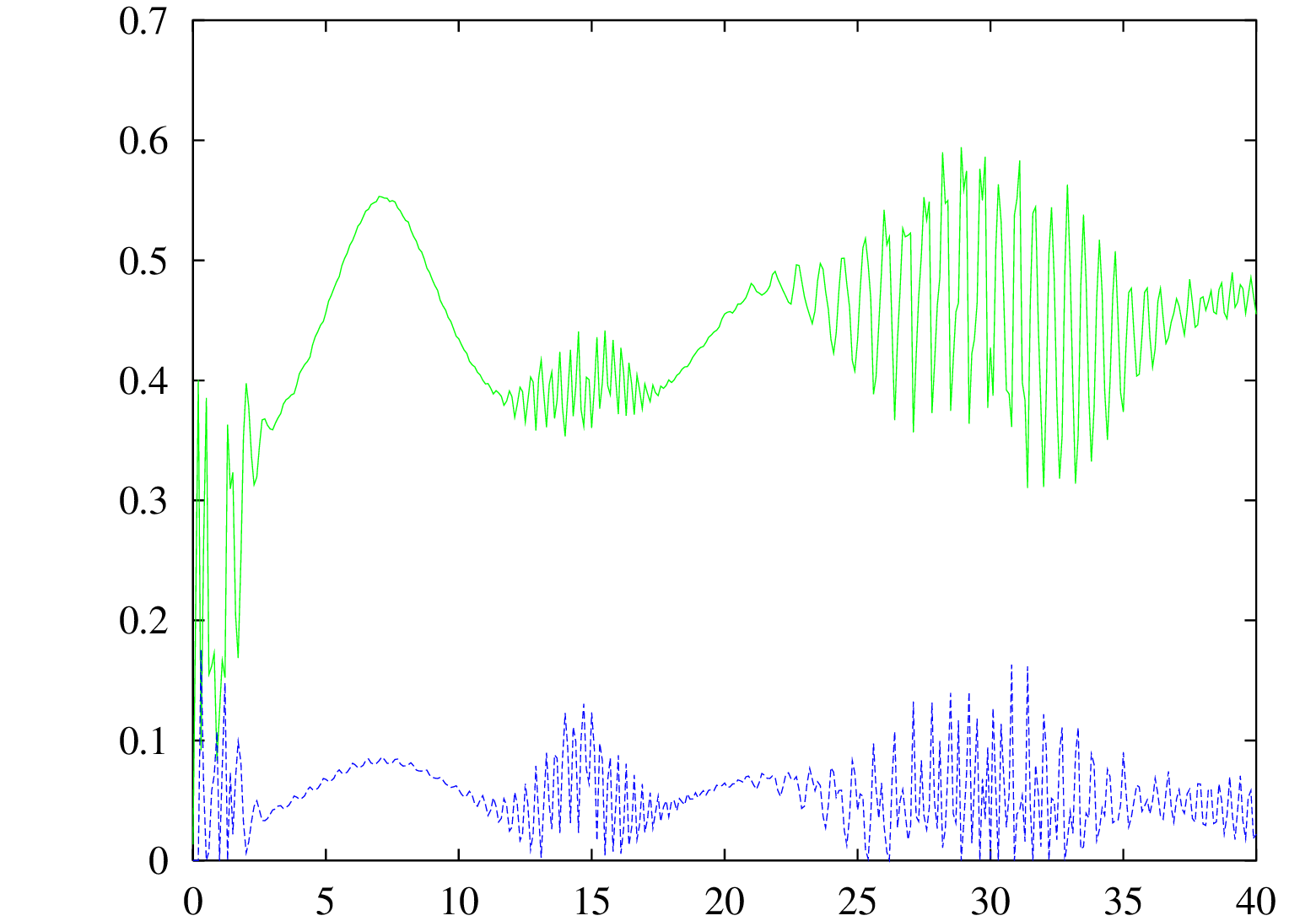}\put(-180,130){ (b) }
\put(-105,-10){ $\lambda t$}
\end{center}
\caption{ The same as Fig. (3), but for the charged qubits system is
initially prepared in the ground state (a) For $\bar{n}=10$ and $\frac{%
\Delta }{\protect\lambda }=0.5.$ (b) For $\bar n=20$ and $\frac{\Delta }{%
\protect\lambda }=1.$ }
\end{figure}

It would be interesting to study more deeply this phenomenon and understand
better its connection to different kinds of entangled states. Hopefully,
this will shed some light on the mysterious nature of multi-party
correlations. Notice that there is a remarkable increases of the classical
deficit $\mathcal{C}_{def}$ for less entangled states. This means that Alice
and Bob can extract more information by means of the $CLOCC$ (classical
communication and local operations). As an example around $\lambda t\simeq 5$%
, the classical correlation is minimum, so the amount of information that
can be extracted in $CLOCC$ by Alice and Bob is minimum, while at $\lambda
t\simeq 10$, where $\mathcal{C}_{c}$ is maximum, the classical deficit is
maximum.

\section{Conclusion}

In conclusion, we have discussed quantum and classical information deficits
of two superconducting charge qubits coupled with a common microwave
resonator. It is shown that, under particular conditions, novel phenomena
can be observed such as a strong quantum correlations, long-lived of
entanglement and entangled states with high degree of entanglement
generation. Moreover the enhancement of quantum correlation is much better
than the classical correlations. Due to fragileness of the entanglement, the
amount of information which has been destroyed in the process of
localization increases for the generated entangled state has a high degree
of entanglement. On the other hand, Alice and Bob can extract more
information by using the local operation and classical communication for
less entangled states. Also, theses results are very important in quantum
cryptography, where if, we consider an eavesdropper(Eve) can make an
additional measurement on the joint state between Alice and Bob, then the
destroyed non-local information will increase for large detuning and small
mean photon number.

\end{document}